# Perspectives for resonances in p-p collisions with the ALICE detector

A. Badalà, G.S. Pappalardo, R. Vernet
*INFN – Sezione di Catania, V. S. Sofia 64, I95123, Catania, ITALY*

F. Blanco, P. La Rocca, C. Petta, A. Pulvirenti, F. Riggi
*Dipartimento di Fisica dell'Università di Catania e INFN-Sezione di Catania, V. S. Sofia 64, I95123, Catania, ITALY*

for the ALICE Collaboration

The capability of the ALICE detector to reconstruct short-lived resonances from the very beginning of the LHC p-p program start-up was investigated, considering about 300,000 minimum bias p-p PYTHIA events at 900 GeV and 10 TeV, fully reconstructed under the hypothesis of a still misaligned detector. Only the information from track reconstruction was taken into account and no PID knowledge was considered. A reliable $\rho(770)$ yield estimation was possible in 900 GeV events. Moreover, in 300,000 p-p events at 10 TeV the signal for the $\phi$-resonance could be extracted. A significance of 9 can be reached if only resonances with a transverse momentum larger than 1.5 GeV/c are considered. Moreover, even with this low statistics the $\phi$ signal is visible in events with a multiplicity as high as 100.

## 1. INTRODUCTION

The Large Hadron Collider (LHC)[1] will start its p-p program with some short runs at 900 GeV, which will permit to test the performance of the LHC detectors. Then, before increasing its energy to the final value of 14 TeV, a rather long acquisition period at 10 TeV is foreseen. Proton-proton collisions will serve both as references for the future nucleus-nucleus collisions (Pb-Pb at 5.5 ATeV) and to test QCD in a new energy domain. Particularly interesting are the high-multiplicity p-p events, for which a particle density ($dN/d\eta \sim$ 50-100) similar to that obtained in mid-central Cu-Cu collisions at RHIC [2] could be reached. Measurement of hadronic resonances is particularly important because these are a powerful probe of the collision dynamics and are useful to understand the hadron formation mechanisms. Moreover, in elementary collisions, strange resonances (such as K* and $\phi$) are sensitive probes of strangeness production, since they have a comparable mass but different quantum numbers.

ALICE [3], the experiment dedicated to relativistic heavy ion physics at LHC, is well suited to measure resonance yields and properties, thanks to its optimal capability to reconstruct and identify electrons with $p_t$ larger than 1 GeV/c and charged particles ($\pi$, K, p) in a quite large $p_t$ range (0.2 GeV/c<$p_t$<5 GeV/c). Further information on the ALICE performance concerning this physics subject can be found in [4,5]. It is interesting to know the possibility for this detector to extract resonances even in a first-day analysis (i.e. with misaligned detectors and without Particle Identification). This possibility has been tested with about 210,000 minimum-bias PYTHIA (6.2) p-p events at 900 GeV and 300,000 minimum-bias PYTHIA p-p events at 10 TeV. Some results for the $\phi$ and $\rho$ resonances are shown in the following paragraphs.

## 2. STUDY OF RESONANCES IN P-P COLLISIONS

Resonances as $\rho(770)$, K*(892), $\phi(1020)$ and $\Lambda(1520)$ will be reconstructed by their prompt decay in $\pi$-$\pi$, $\pi$-K, K-K and K-p. The central part of the ALICE detector (the TPC and the Inner Tracking System) covers the pseudo-rapidity range $|\eta| < 0.9$ and is able to track charged particles with a $p_t$>0.1 GeV/c with a momentum resolution better then 0.7% for $p_t$<2 GeV/c and better then 3 % for $p_t$< 100 GeV/c. Such detectors, combined with the Time Of Flight (TOF), are able to identify pions, kaons and protons in the range 0.1 GeV/c<$p_t$<4 GeV/c, after a dedicated dE/dx and time-of-flight calibration. Using cosmics and first p-p data, an overall alignment procedure will be carried out in order to have a residual misalignment of about 10 μm between the different parts of the ITS detector and of about 100 μm between the TPC and the ITS.

However, to study the ALICE possibility to extract physical information just from the beginning of its data taking, proton-proton events generated by PYTHIA were injected in a detector software replica in which the maximum expected misalignments (i.e as expected before any alignment procedure) between the different parts of the Inner Tracking System detector (about 100 μm) and between the TPC and the ITS (2 mm) were introduced.

## 2.1. p-p collisions at 900 GeV

About 210,000 p-p minimum bias PYTHIA events at 900 GeV were used to extract the hadronic resonance signals. If only the TPC information is taken into account, without any PID information, it is possible to extract with a reliable confidence the ρ(770) peak at least in some $p_t$ bins. An example of the ρ(770) signal extracted in the range $p_t$<1 GeV/c is shown in Fig.1. In this case the difference between reconstructed and true signal yield is less than 5%, for a peak with a significance of about 30.

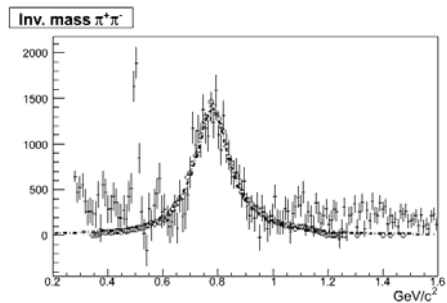

Figure 1: ρ(770) signal for $p_t$<1 GeV/c. The $k_s^0$ peak [5] is also visible at about 0.5 GeV/$c^2$. The true ρ signal (open points) and the extracted signal estimated by a Breit-Wigner fit (dashed-dotted line) are also shown.

## 2.2. p-p collisions at 10 TeV

About 300,000 p-p minimum bias PYTHIA events at 10 TeV were fully reconstructed. Extraction of the φ(1020) signal without any PID information was done in different $p_t$ bins and for different $p_t$ cuts. The background was estimated by the event mixing technique and by the like-sign method. However, the best results were obtained when the background was fitted by a 3[rd] order polynomial function and the signal by a Breit-Wigner. The best significance (about 9) is obtained if a cut on the $p_t$ of the $K^+K^-$ pair selecting $p_t$>1.5 GeV/c is applied. With this cut, the signal can be extracted for different charged particle multiplicity bins even if the significance for high multiplicity events is rather low (S/√S+B <3) (see fig.2). A comparison of extracted and true yields as a function of the charged particle multiplicity is reported in fig.3. A good agreement, except for the really low multiplicity bin, is achieved.

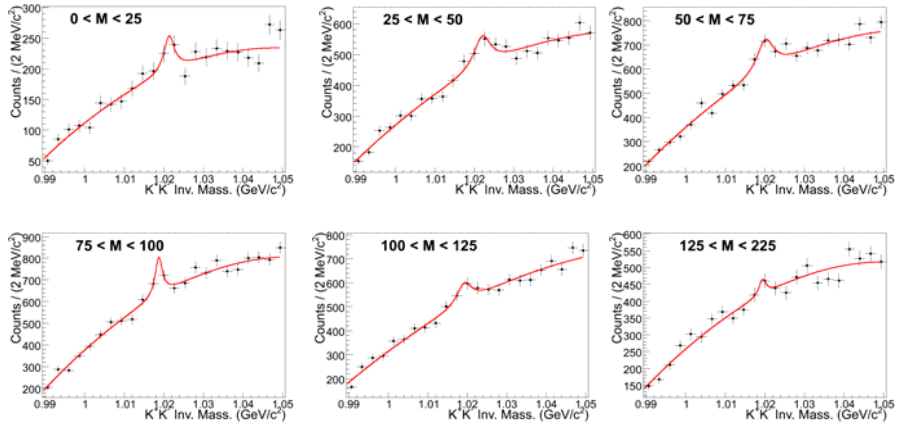

Figure2: φ(1020) signal for different charged particle multiplicity bins

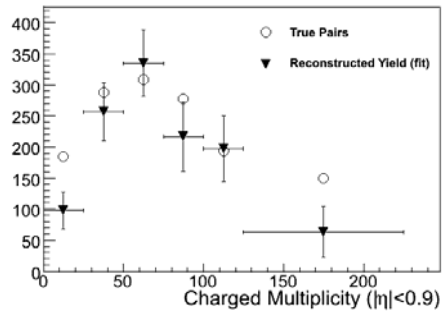

Figure 3: True and reconstructed φ(1020) as a function of charged particle multiplicity

## 3. CONCLUSION

Studies on the extraction of the ρ(770) and φ(1020) signal in the case of a fully misaligned ALICE detector and with only tracking information were done. About 200,000 and 300,000 minimum bias PYTHIA p-p events at the collision energy of 900 GeV and 10 TeV were analyzed. Reliable extraction of these resonances was shown to be possible. Particularly important is the extraction of the φ peak also in high multiplicity events. However, an estimation of the significance as a function of $p_t$ and event multiplicity shows that at least 1M of events is needed to have a rather large significance (> 10) in a large $p_t$ and multiplicity range.